\documentclass[a4paper,12pt]{article}
\usepackage{amsmath,amsfonts,amssymb,epsfig,subfigure, cite}

\renewcommand{\le}{\leqslant}
\renewcommand{\ge}{\geqslant}

\newcommand{\be}{\begin{equation}}
\newcommand{\en}{\end{equation}}

\newcommand{\ii}{\textrm{i}}
\newcommand{\ee}{\textrm{e}}
\renewcommand{\vec}[1]{\boldsymbol{#1}}

\begin{document}

\title{Surface Instability of Sheared Soft Tissues}
\author{
  M.~Destrade,   M.D.~Gilchrist,   \\D.A.~Prikazchikov,   G.~Saccomandi}
\date{2008}
\maketitle


\begin{abstract}

When a block made of an elastomer is subjected to large shear,
its surface remains flat.
When a block of biological soft tissue is subjected to large shear,
it is likely that its surface in the plane of shear will buckle
(apparition of wrinkles).
One factor that distinguishes soft tissues from
rubber-like solids is the presence -- sometimes visible to the naked eye --
of oriented collagen fibre bundles, which are stiffer than the elastin matrix 
into which they are embedded but are nonetheless flexible and extensible.
Here we show that the simplest model of isotropic nonlinear elasticity,
namely the incompressible neo-Hookean model, suffers surface instability in shear 
only at tremendous amounts of shear, i.e., above 3.09, which corresponds to a $72^\circ$
angle of shear. 
Next we incorporate a family of parallel fibres in the model and show that the resulting 
solid can be either reinforced or strongly weakened with respect to surface instability,
depending on the angle between the fibres and the direction of shear,
and depending on the ratio $E/\mu$ between the stiffness of the fibres and that of the matrix. 
For this ratio we use values compatible with experimental data on soft tissues. 
Broadly speaking, we find that the surface becomes rapidly unstable when the shear takes place 
``against'' the fibres, and that as $E/\mu$ increases, so does the sector of angles where 
early instability is expected to occur. 

\end{abstract}



\section{Introduction}

\begin{figure}
\centering 
\epsfig{figure=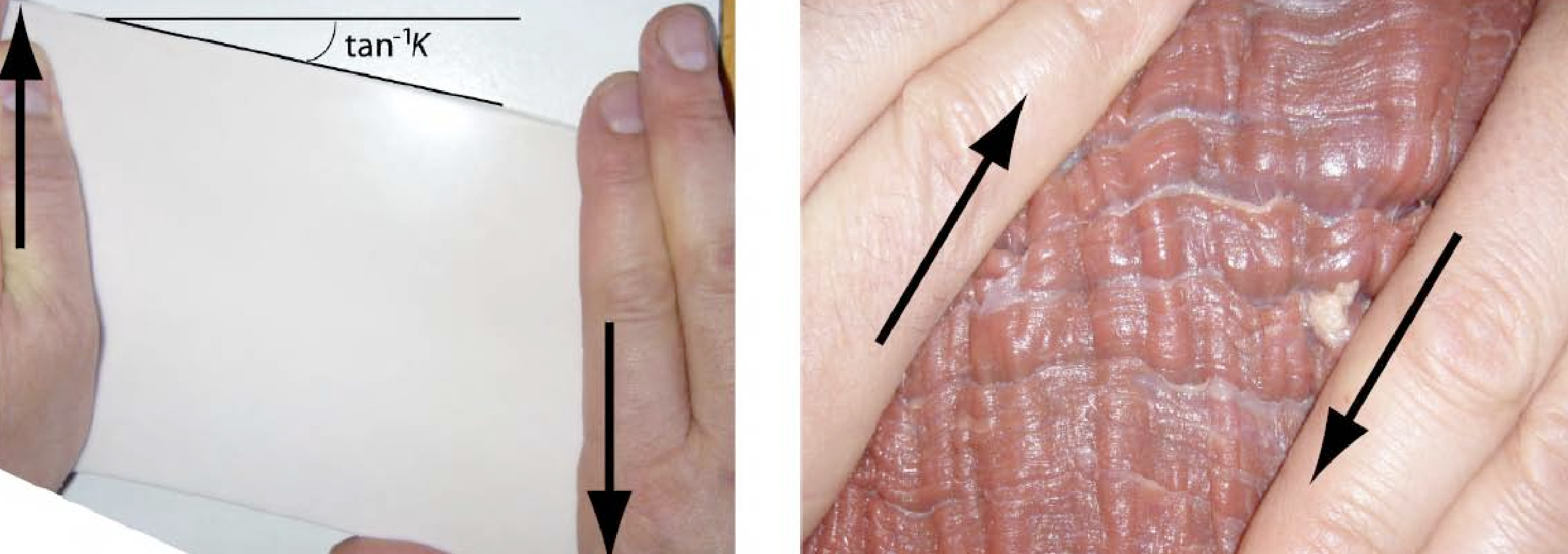, width=.95\textwidth, height=.35\textwidth}
 \caption{Shearing (along the arrows) a block of silicone of approximate size 
 $15 \text{cm} \times 10 \text{cm} \times 1.5 \text{cm}$ 
 and a block of mammalian skeletal muscle (beef) of approximate size 
  $15 \text{cm} \times 10 \text{cm} \times 3 \text{cm}$;
  one does not exhibit surface instability, the other does.} 
\label{fig_silicone_meat}
\end{figure}

Rubber-like solids and biological soft tissues can both be efficiently 
modelled within the framework of finite elasticity, which can 
account for large deformations, physical nonlinearities, incompressibility, 
residual stresses, viscoelasticity, etc. 
One of the most salient differences between the two types of solids is 
that at rest, elastomers are essentially isotropic whilst soft
tissues are essentially anisotropic, because of the presence of 
collagen fibre bundles. 
In that respect, it is worthwhile to consider the effect of
incorporating 
families of parallel fibres into an isotropic matrix, and see if it can
model some striking differences between the 
mechanical behaviour of elastomers and of soft tissues.
Consider for instance the large shear of a solid block.
When the block is made of an elastomer such as silicone, its surface 
remains stable;
when it is made of a biological soft tissue such as skeletal muscle, 
its surface wrinkles for certain ranges of orientation between the 
direction of shear and the (presumed) direction of fibres, see Fig.~\ref{fig_silicone_meat}.
Here we show that one of the simplest models of anisotropic nonlinear elasticity, 
which requires only knowledge of the fibre/matrix stiffness ratio, 
is sufficient to successfully predict these behaviours.

To model the isotropic elastomer (Section 2), we take the incompressible neo-Hookean 
solid, and find that it does not suffer surface instability unless it is subjected 
to a substantial amount of shear (critical amount of shear: 3.09, critical angle of shear: $72^\circ$).
In that case the wrinkles are aligned with the direction of greatest stretch.
(The wrinkling analysis relies on the incremental theory of nonlinear
elasticity, see for instance Biot \cite{Biot63} or Ogden
\cite{Ogde84}).
Next, we introduce one family of parallel fibres into the model (Section 3).
To model biological soft tissues with one preferred direction  (Section 4), we take 
the incompressible neo-Hookean strain energy density, augmented by the so-called 
`standard reinforcing model':
this model has only two parameters, namely the shear modulus $\mu$ of the soft (neo-Hookean)
matrix and the fibre stiffness $E$.
 
With respect to surface instability, only the ratio $E/\mu$ of these two quantities plays a role.
We take it to be equal in turn to $40.0$, $20.0$, and $10.0$, 
in agreement with the range of experimental measures found in the literature. 
We then find that when the angle between the direction of shear and the direction of the fibres is
small, the solid is much more stable than the isotropic solid obtained in the absence of fibres;
when the angle increases but is less than $99.0^\circ$ (for $E/\mu = 40.0$), 
$102.8^\circ$ (for $E/\mu = 20.0$), $108.1^\circ$ (for $E/\mu = 10.0$),  the solid 
remains more stable than the isotropic solid;
however, when the angle exceeds those values,
the critical amount of shear for surface instability 
drops to extremely low levels, indicating the appearance of wrinkles as
soon as shearing occurs.
In that case, the wrinkles are found to be almost orthogonal to the
fibres, in accordance with visual observations.

It is hoped that the paper provides a 
greater understanding of the causes of certain instabilities 
in soft tissues and a quantitative tool 
to measure what deformations (critical amounts of shear) are permissible 
and in which directions. 
Surface instability has a direct connection to slab and tube 
buckling, which in biomechanics may potentially translate into 
aneurysms formation, arterial kinking and tortuosity, brain trauma, and 
many other, still not well understood, pathologies.  


\section{Surface instability of a sheared isotropic solid}


First, we recall known results in the theory of surface wrinkling 
valid for \emph{isotropic} solids.

Consider a semi-infinite body made of an incompressible isotropic neo-Hookean solid, 
for which the strain energy function $W$, written as a function of 
the principal stretch ratios $\lambda_1$, $\lambda_2$, $\lambda_3$, 
is given by
\be
W = \mu (\lambda_1^2 + \lambda_2^2 + \lambda_3^2 - 3)/2.
\en
Here $\mu$ is the shear modulus, and $\lambda_1 \lambda_2 \lambda_3 =1$
by the incompressibility constraint.
Then subject the solid to a large homogeneous static deformation, such that 
$\lambda_2$ is the stretch ratio along the normal to the free surface.
It has long been known that the surface becomes unstable when 
the following \emph{wrinkling condition} is met,
\be \label{wrinkling_neo}
\lambda_1^2 \lambda_3 = \sigma_0,
\en
where $\sigma_0 \simeq 0.296$ is the real root of 
$\sigma^3 + \sigma^2 + 3\sigma - 1 =0$
(Green and Zerna \cite{GrZe54}, Biot \cite{Biot63}).

\begin{figure}
\centering \mbox{\subfigure{\epsfig{figure=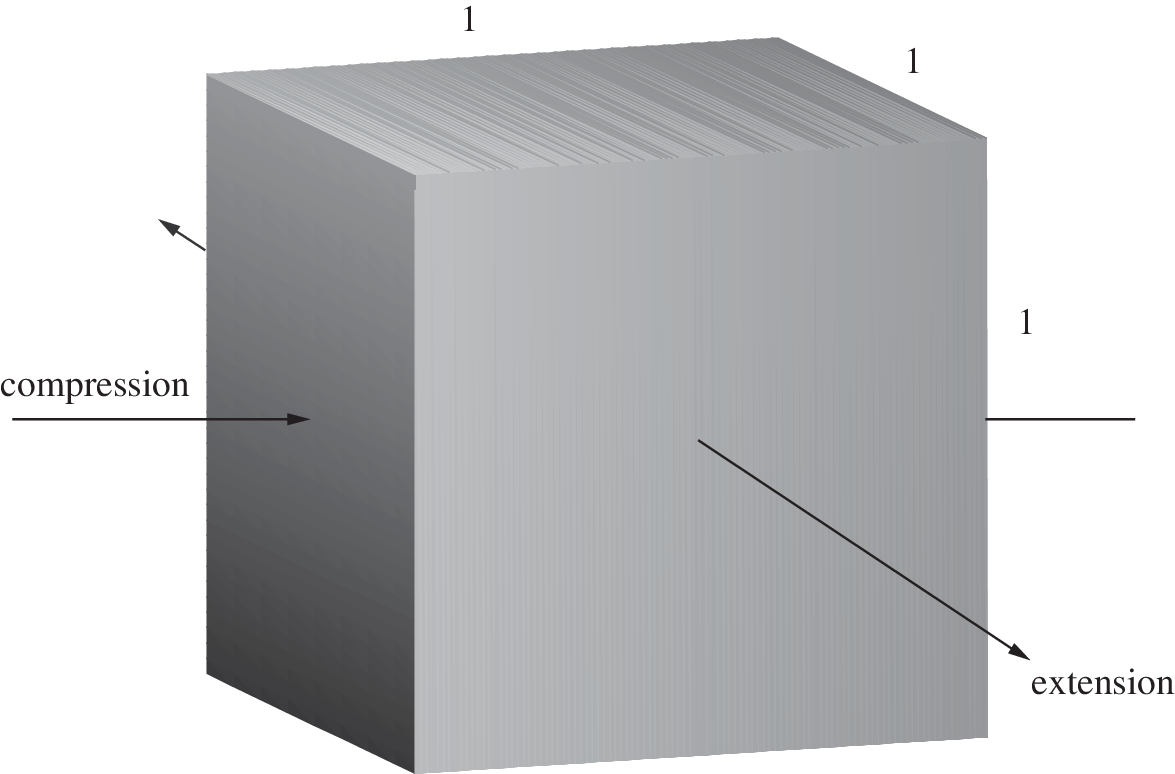,
width=.53\textwidth}}}
  \quad \quad
     \subfigure{\epsfig{figure=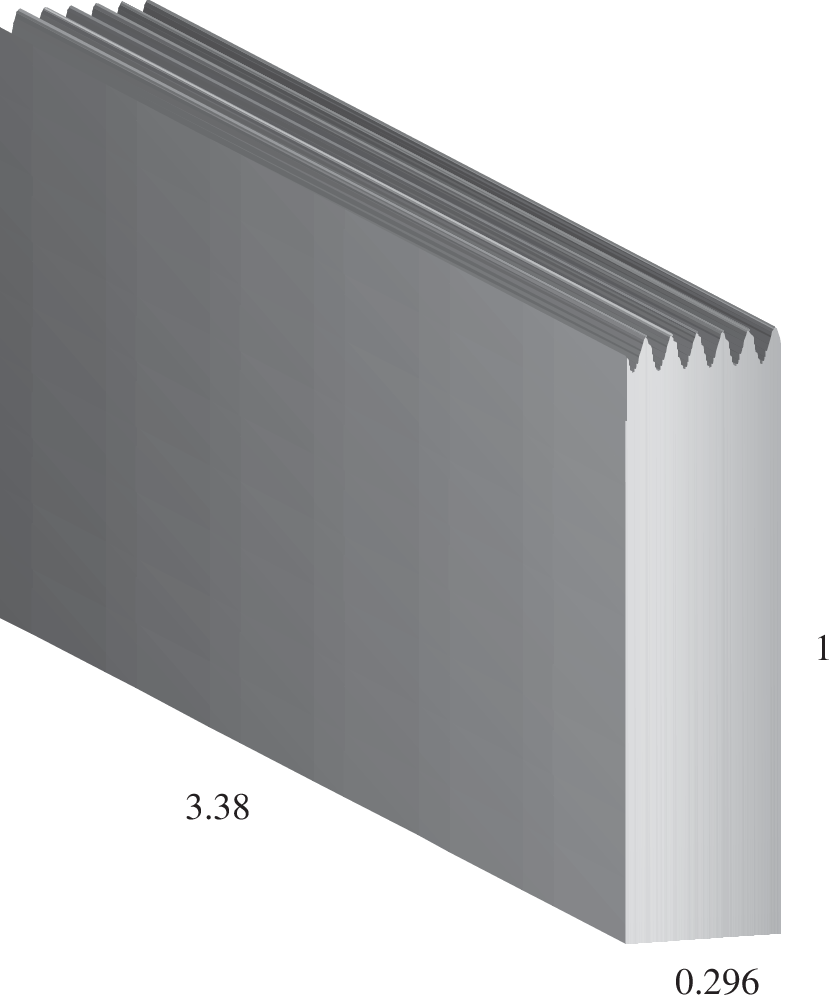,
width=.39\textwidth}}
 \caption{Large plane strain deformation of a unit cube 
 near the surface of a semi-infinite incompressible neo-Hookean solid. 
 When the solid is compressed by 71\%
 (or equivalently, stretched by 238\%), its
 surface wrinkles. Note that the analysis quantifies neither the
amplitude nor wavelength of the wrinkles.}
 \label{compress_isotropic}
\end{figure}

In the following \emph{plane strain} situation,
\be \label{plane_strain}
 \lambda_1 = \lambda, \qquad \lambda_2 = 1, \qquad \lambda_3 = \lambda^{-1},
\en
the critical stretch of compression found from Eq.~\eqref{wrinkling_neo}
is clearly $\lambda_1 = \sigma_0 \simeq 0.296$ 
(and then $\lambda_3 = \sigma_0^{-1} \simeq 3.38$). 
The conclusion is that when a semi-infinite neo-Hookean solid, which is neither allowed to 
expand nor contract along the normal to its boundary, 
is \emph{compressed} by 71\% in a given direction (lying in the 
boundary), it buckles with wrinkles developing along the direction 
\emph{orthogonal} to the direction of compression.
Equivalently, when it is \emph{stretched} by 238\%, it 
buckles with wrinkles \emph{parallel} to the direction of tension. 
Figure~\ref{compress_isotropic} summarizes these results. 

It is natural to wonder whether the surface might have become unstable 
in other directions earlier, that is at compressive ($\le 1$) ratios 
larger than 0.296, or at tensile ($\ge 1$) ratios smaller than 3.38.
Flavin \cite{Flav63} shows that wrinkles develop parallel to the
direction 
making an angle $\theta$ with the principal direction of strain associated 
with the stretch ratio $\lambda_3$ when the following wrinkling 
condition is met
\be 
\lambda_1^2 \lambda_3^2 ( \lambda_1^2 \cos^2\theta + 
  \lambda_3^2 \sin^2\theta) = \sigma_0^2.
\en
In the plane strain situation Eq.~\eqref{plane_strain}, this condition 
is quadratic in $\lambda^2$,
\be \label{biquadratic}
\lambda^4 \cos^2\theta - \lambda^2 \sigma_0^2 + \sin^2\theta = 0.
\en
It has real roots provided $\theta$ is in the ranges 
$-\theta_0 \le \theta \le \theta_0$ or 
$\pi/2 - \theta_0 \le \theta \le \pi/2 + \theta_0$, 
where $\theta_0 = (1/2)\sin^{-1} \sigma_0^2 \simeq 2.51^\circ$.
In the former range, the compressive critical stretch found from 
the biquadratic Eq.~\eqref{biquadratic} turns out to be smaller than 
0.296 and in the latter range, to be larger than 3.38.
Thus surface instability for plane  
strain Eq.~\eqref{plane_strain} occurs when the isotropic neo-Hookean
half-space is in 
compression at a ratio $\sigma_0$ or equivalently, in tension at a ratio $\sigma_0^{-1}$.
The \emph{wrinkles are parallel to the direction of greatest stretch} and orthogonal 
to the direction of greatest compression.

\begin{figure}
\centering 
\epsfig{figure=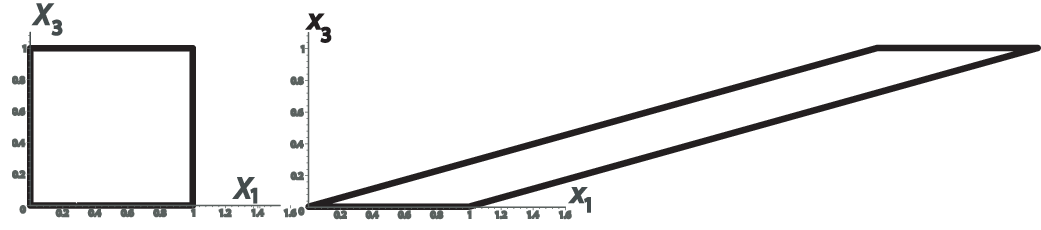, width=.9\textwidth}
 \caption{
 Large simple shear of a unit square in 
 the surface of a semi-infinite incompressible neo-Hookean solid. 
 When the solid is sheared by an amount $K_0 \simeq 3.09$
 (Figure on the right), its
 surface wrinkles. The corresponding angle of shear is 
 $\tan^{-1} K_0 \simeq 72.0^\circ$, which physically, is abnormaly
large.
 Then the wrinkles are parallel to the 
direction of greatest tension, which makes an angle 
$\varphi_0  \simeq 16.5^\circ$ with the direction of shear (and so,  
the wrinkles are almost aligned with the sheared faces.) }
 \label{fig_shear_isotropic}
\end{figure}

Now \emph{simple shear} belongs to the family of plane strains
Eq.~\eqref{plane_strain}, with the following connection 
between the principal stretches and the amount of shear $K$ 
(see Ogden \cite{Ogde84}) for instance),
\be \label{shear}
K = \lambda - \lambda^{-1}, \qquad 
\lambda = K/2 + \sqrt{1+K^2/4}.
\en
Also, the \emph{direction of greatest stretch} is at an angle $\psi$ with the direction 
of shear, where $\psi \in ]0, \pi/4]$ is given by
\be \label{greatest_stretch}
\tan 2 \psi = 2/K.
\en
Clearly $\lambda>1$ here, and so surface shear instability occurs in 
tension, when the amount of shear is equal to 
$K_0 = \sigma_0^{-1} - \sigma_0 \simeq 3.09$.
The corresponding critical angle of shear is then 
$\tan^{-1} K_0 \simeq 72.0^\circ$,
see Fig.~\ref{fig_shear_isotropic}.
This is quite large shear. 

\section{Sheared fibre-re\-in\-for\-ced solids}



\subsection{Finite simple shear}


Now we consider a semi-infinite composite incompressible solid, made of 
an iso\-tro\-pic matrix reinforced with one family of parallel extensible fibres,
themselves parallel to the boundary of the solid.
In the undeformed configuration, we call $( X_1, X_2, X_3)$ 
the set of Cartesian coordinates
such that the solid is located in the $X_2 \ge 0$ region.
We denote by $\vec{E}_1$,  $\vec{E}_2$,  $\vec{E}_3$ the orthogonal unit 
vectors defining the Lagrangian (reference) axes, aligned with the 
$X_1$, $X_2$, $X_3$ directions, respectively. 

When the solid is sheared in the direction of $\vec{E}_1$, 
the particle at $\vec{X}$ moves to its
current position $\vec{x}$. 
We call $\vec{F} = \partial \vec{x}/\partial \vec{X}$ the associated
deformation gradient tensor, and $\vec{B}= \vec{F} \vec{F}^T$ the left Cauchy-Green 
strain tensor. 
We then call ($x_1, x_2, x_3$) the Cartesian coordinates, aligned 
with ($X_1, X_2, X_3$), corresponding to the current position
$\vec{x}$.
In the current configuration, the basis vectors are $\vec{e}_1$, 
$\vec{e}_2$, $\vec{e}_3$, and here they are such that 
$\vec{e}_i \equiv \vec{E}_i$ ($i=1,2,3)$. 
The \emph{simple shear} of amount $K$ is described by
\be \label{shear_deformation}
x_1 = X_1 + K X_3, \qquad x_2 = X_2, \qquad x_3 = X_3.
\en
We thus find in turn that 
\be \label{B}
\vec{F} = \vec{I} + K \vec{e}_1 \otimes \vec{E}_3,
\qquad
 \vec{B} = \vec{I} + K(\vec{e}_1 \otimes \vec{e}_3 
  + \vec{e}_1 \otimes \vec{e}_3)  + K^2 \vec{e}_1 \otimes \vec{e}_1. 
\en
The principal stretches are given by Eq.~\eqref{plane_strain} 
and Eq.~\eqref{shear}, and the first principal isotropic invariant 
$I_1 = \text{tr }\vec{B}$ is given here by 
\be
I_1 = 3 + K^2.
\en
Note that for shear, the second principal isotropic invariant, 
$I_2 = [I_1^2 - \text{tr }(\vec{B}^2)]/2$ is also equal to 
$3+K^2$.


\subsection{One family of fibres}


For solids reinforced with one family of parallel fibres
lying in the plane of shear, we work in 
all generality and consider that the angle $\Phi$ (say) between the fibres and 
the $X_1$ direction can take any value.  
In other words, the unit vector $\vec{M}$ (say)
in the preferred fibre direction has components 
\be \label{M}
\vec{M} = \cos \Phi \vec{E}_1 + \sin \Phi \vec{E}_3, 
\en
in the reference configuration.
Simple shear is a homogeneous deformation, and so $\vec{M}$ 
is transformed into $\vec{m} = \vec{F M}$ in the current configuration, 
that is
\be \label{m}
\vec{m} = (\cos \Phi + K \sin \Phi)\vec{e}_1 + \sin \Phi \vec{e}_3.
\en
Without loss of generality, we take the ranges
$K \ge 0$, $0 \le \Phi \le \pi$,
which cover all possible orientations of the fibres with respect to 
the direction of shear.

\begin{figure}
\centering 
\epsfig{figure=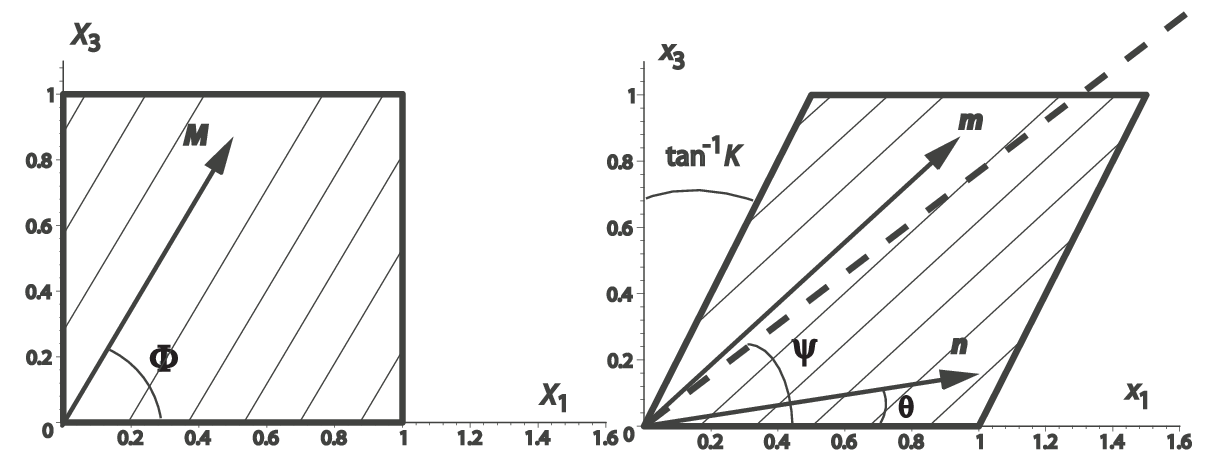, width=.9\textwidth}
 \caption{
 A unit square lying on the surface of a semi-infinite solid reinforced 
 with one family of fibres (thin lines) and subject to a simple shear of amount $K = 0.5$
 (angle of shear is $\tan^{-1} K \simeq 26.6^\circ$) in the $X_1$ direction.
 In the reference configuration, the fibres are along the unit vector $\vec{M}$,
 at the angle $\Phi = 60^\circ$ with the $X_1$-axis.  
 In the current configuration, they are along $\vec{m}$.
 The unit vector $\vec{n}$ is orthogonal to the wrinkles' front (when they exist).
 Finally, the dashed line is aligned with the direction of greatest stretch;
 it is at an angle $\psi \simeq 38^\circ$ to the direction of shear.}
 \label{fig_one_family}
\end{figure}

To fix the ideas, consider Fig.~\ref{fig_one_family}.
There we shear the half-space by a finite amount $K = 0.5$,
in the direction making an angle $\Phi = 60^\circ$ with the fibres. 
Notice that a unit vector $\vec{n}$ 
making an angle $\theta$ with the direction of shear is also 
represented in the current configuration.
This is the normal to the wrinkles' front; 
in the next section we look for surface wrinkles in all directions
(the angle $\theta$ spans the interval $[0^\circ, 180 ^\circ]$) 
and we determine which is the smallest corresponding critical amount of shear.

Finally we introduce the anisotropic invariants $I_4 \equiv \vec{m \cdot m}$
and $I_5 \equiv \vec{F m \cdot}\vec{F m}$;
in particular we find 
\be \label{I4}
 I_4 = 1 + K \sin 2 \Phi + K^2 \sin^2 \Phi.
\en
Recall that $I_4$ is the squared stretch in the fibre direction 
\cite{Spen84}. 
In particular, if $I_4 \ge 1$ then the fibres are in extension,
and if $I_4 \le 1$ then they are in compression. 
Clearly here, when $0 \le \Phi \le \pi/2$, the fibres are always in
extension but when $\pi/2 < \Phi < \pi$, there 
exist a certain amount of shear (explicitly, $-2 /\tan \Phi$)
below which the fibres are in compression.


\subsection{Constitutive assumptions}


In general, the strain-energy density $W$ of a hyperelastic incompressible
solid reinforced with one family of parallel extensible fibres depends on 
the isotropic invariants $I_1$ and $I_2$, and on the
anisotropic invariants (Spencer \cite{Spen84}) $I_4$ and $I_5$.
We assume that $W$ is the sum of 
an isotropic part and an anisotropic part. 
For the isotropic part, modelling the properties of the `soft' matrix, 
we take the neo-Hookean strain-energy density in order to make 
a connection with the results of Section 2.
For the anisotropic part, modelling the properties of the extensible `stiff' 
fibres, we take a function of $I_4$ only, say $F(I_4)$.
Hence, we restrict our attention to those solids with strain energy density
\be \label{W}
W = \mu(I_1 - 3)/2 + F(I_4).
\en
This assumption is quite common in the biomechanics literature. 
Although it does not prove crucial to the analysis, it leads to compact and revealing expressions
(Note that the consideration of a more general $W$ poses no major extra difficulty,
but results in much longer expressions.). 

The corresponding Cauchy stress tensor $\vec{\sigma}$ is (see e.g.
\cite{Ogde03}):
$\vec{\sigma} =  - p \vec{I} + \mu \vec{B} 
 + 2 F'(I_4) \vec{m} \otimes \vec{m}$, 
where $p$ is a Lagrange multiplier introduced by the constraint
of incompressibility.
The surface $x_2 = 0$ is free of tractions:
here $\sigma_{12} = \sigma_{23} = 0$ follows from $B_{12} = B_{23} = 0$
and $\vec{m \cdot e}_2 = 0$
(see Eq.~\eqref{B} and Eq.~\eqref{m}), 
whilst $\sigma_{22} = 0$ gives $p = \mu$. 
Thus, the pre-stress 
necessary to maintain the shear Eq.~\eqref{shear_deformation} is 
\be
\vec{\sigma} =  \mu (\vec{B}  - \vec{I})
 + 2 F'(I_4) \vec{m} \otimes \vec{m},
\en 
showing that the directions of principal stress and  
strain do not coincide in general (except when the
preferred direction
is aligned with principal directions of strain).


\section{Surface instability}



\subsection{Incremental deformations}


We seek solutions to the incremental equations of equilibrium 
and incremental boundary conditions in the form of a 
sinusoidal perturbations whose amplitude 
decays rapidly with depth. 
In contrast to the isotropic case of Section 2, 
we do not know a priori in which direction the wrinkles should be
aligned, and we take the normal to the wrinkles' front 
$\vec{n}$ (say) to lie in the ($x_1 x_3$) plane at an arbitrary 
angle $\theta$ with $x_1$, see Fig.~\ref{fig_one_family}.
Hence, we seek a perturbation solution $\vec{u}$ (mechanical displacement)
and $\dot{p}$ (increment of the Lagrange multiplier associated with 
incompressibility) in the form,
\begin{equation} \label{increment}
\{ \vec{u}, \dot{p} \}
  = \{ \vec{U}(k x_2), \ii k P(k x_2)\}
                \ee^{\ii k(\cos\theta x_1 + \sin \theta x_3)},
\end{equation}
where $k$ is the ``wave''-number and $\vec{U}$, $P$ are functions of $k
x_2$ alone.

The incremental equations read
\begin{equation} \label{equilibrium}
s_{j i,j} = 0, \qquad u_{j,j} = 0,
\end{equation}
where the comma denotes partial differentiation with respect to $x_j$,
and $\vec{s}$ is the incremental nominal stress tensor. 
Its components are \cite{Ogde84},
\begin{equation} \label{s}
s_{j i} = \mathcal{A}_{0jilk}u_{k,l} + p u_{j,i} - \dot{p} \delta_{i j},
\end{equation}
where $\vec{\mathcal{A}_0}$ is the fourth-order tensor of instantaneous 
elastic moduli.
In general it has a long expression for fibre-reinforced solids, with possibly
45 non-zero components, see for example \cite{ChWh86, PrRo04}.
For $W$ in the form Eq.~\eqref{W}, $\vec{B}$ by Eq.~\eqref{B},
and $\vec{M}$ by Eq.~\eqref{M},  we find the following components 
\begin{equation} \label{moduli}
\mathcal{A}_{0jilk} = \mu \delta_{ik} B_{jl}
 + 2 F'(I_4) \delta_{ik} m_j m_l
   + 4 F''(I_4) m_i m_j m_k m_l
\end{equation}
see Merodio and Ogden \cite{MeOg02}. 
Clearly, these components have the symmetries
$\mathcal{A}_{0jilk} = \mathcal{A}_{0lkji}$
and $\mathcal{A}_{0jilk} = \mathcal{A}_{0jkli}$. 
We end up with 23 non-zero components, 
several of which are equal to one another
(\emph{in toto} there are 13 different components). 

Clearly, if $\vec{u}$ and $\dot{p}$ are of the form Eq.~\eqref{increment},
then by Eq.~\eqref{s} the $s_{j i}$ are of a similar form, say
\begin{equation} 
s_{j i}
  =  \ii k S_{j i}(k x_2)
                \ee^{\ii k(\cos\theta x_1 + \sin \theta x_3)},
\end{equation}
where the $S_{j i}$ are functions of the variable $k x_2$ only.
By a systematic procedure, first laid down by Chadwick \cite{Chad97}
(see also \cite{DeOg05, DOPR05, Fu05a, Fu05b}),
we can eliminate $P$ and write the incremental equations of equilibrium 
as a first-order differential system. 
This is known as the \emph{Stroh formulation} of the problem,
\begin{equation} \label{1stOrder}
  \begin{bmatrix}
     \vec{U}' \\ \vec{S}' 
  \end{bmatrix} 
    = \ii \vec{N} \begin{bmatrix}
                          \vec{U} \\ \vec{S} 
                         \end{bmatrix},
    \quad \text{where} \quad
  \vec{U} = \begin{bmatrix} U_1 \\ U_2 \\ U_3 \end{bmatrix}, \quad 
  \vec{S} = \begin{bmatrix} S_{21} \\ S_{22} \\ S_{23} \end{bmatrix}, \quad
  \vec{N} = \begin{bmatrix}
             \vec{N}_1 & \vec{N}_2 \\
             \vec{N}_3 & \vec{N}_1
           \end{bmatrix},
\end{equation}
and the symmetric $3 \times 3$ matrices $\vec{N}_1$, $\vec{N}_2$, $\vec{N}_3$ are 
given by
\begin{equation}  \label{N1N2N3}
- \vec{N}_1 =
      \begin{bmatrix}
       0 & \cos \theta & 0 \\
       \cos \theta & 0 & \sin \theta \\
       0 & \sin \theta & 0
       \end{bmatrix},
\quad
 \vec{N}_2 =
  \begin{bmatrix}
       1/\mu & 0 & 0 \\
           0 & 0 & 0 \\
        0 & 0 &  1/\mu
      \end{bmatrix},
\quad
 - \vec{N}_3 = \begin{bmatrix}
                       \eta   &     0     & \kappa \\
                          0   &     \nu   &      0     \\
                     \kappa  &     0     &  \chi
                   \end{bmatrix},
\end{equation}
with
\begin{align}
& \eta =
    (\mathcal{A}_{01111} + 3 \mu) \cos^2 \theta 
     + 2 \mathcal{A}_{01131} \cos \theta \sin \theta 
          + \mathcal{A}_{03131} \sin^2 \theta,
\nonumber \\
& \nu =
   \mathcal{A}_{01212} \cos^2 \theta 
     + 2 \mathcal{A}_{01232} \cos \theta \sin \theta 
          + \mathcal{A}_{03232} \sin^2 \theta - \mu,
\nonumber \\
& \chi =
     \mathcal{A}_{01313} \cos^2 \theta 
     + 2 \mathcal{A}_{01333} \cos \theta \sin \theta 
          + (\mathcal{A}_{03333} + 3 \mu)\sin^2 \theta,
\nonumber \\
& \kappa =  
    \mathcal{A}_{01113} \cos^2 \theta 
     + (2 \mathcal{A}_{01133}  + 3 \mu)\cos \theta \sin \theta 
          + \mathcal{A}_{03133} \sin^2 \theta.
\end{align}
Notice how  all the information relative to anisotropy is located in the $\vec{N}_3$ matrix.

The solution to the system Eq.~\eqref{1stOrder} is clearly an exponential
\be
\{ \vec{U}, \vec{S}\} = \{ \vec{U}^0, \vec{S}^0\} \ee^{\ii k q x_2},
\en 
where $\vec{U}^0, \vec{S}^0$ are constant vectors and $q$ is an eigenvalue of $\vec{N}$.
The characteristic equation associated with $\vec{N}$ is a
\emph{bicubic} \cite{PrRo04},
\be \label{bicubic}
q^6 - \left(2 - \frac{\chi + \eta}{\mu} \right) q^4
 + \left(1 + \frac{\nu - 2\epsilon}{\mu}  + \frac{\chi \eta - \kappa^2}{\mu^2} \right) q^2
  + \frac{\epsilon (\mu + \nu)}{\mu^2} = 0,
\en
where the quantity $\epsilon$ is defined by
\be
\epsilon = 
\chi \cos^2 \theta 
     - 2 \kappa \cos \theta \sin \theta 
          + \eta \sin^2 \theta.
\en
The existence of real roots to this equation corresponds to the loss
of ellipticity of the governing equations (\emph{material} instabilities). 
This possibility has been thoroughly investigated before, see
\cite{TrAb83, QiPe97, MeOg02}.
Here we focus on complex roots and keep those satisfying 
$\text{Im} q > 0$ , for a surface-type bifurcation which decays 
with depth (\emph{geometric} instability).


\subsection{Wrinkling condition and resolution scheme}


Over the years, many schemes have been developed to solve 
surface boundary problems using the Stroh formulation;
we used in turn the determinantal 
method \cite{Farn70}, 
the Riccati matrix equation of surface impedance \cite{Fu05a,
Fu05b},
and explicit polynomial equations
\cite{Dest05}, in order to double-check 
our numerical computations.

The crucial boundary condition is to find the amount of shear at which the 
surface of the sheared solid is free of tractions. 
The safest way to express this is
\be \label{detM}
\text{det } \vec{Z} =0,
\en
where $\vec{Z}$ is the (Hermitian) \emph{surface impedance matrix},
which relates tractions to displacements through $\vec{S} = \ii \vec{Z U}$. 
We remark that the schemes are not as safe in surface stability problems 
as they are in surface wave theory because of incompressibility
\cite{Fu05a, Fu05b}
and non-monotonicity of $\text{det } \vec{Z}$ with $K$.

Once Eq.~\eqref{detM} is reached, 
we can construct an incremental solution to the equations
of equilibrium which is adjacent to the large shear 
equilibrium, and signals the onset of surface instability.
We adopted the following strategy:
\begin{enumerate}
 \item Fix $\Phi$, the angle between the direction of shear and the preferred 
 direction;
 \item Fix $\theta$, the angle between the direction of shear and the 
 normal to the wrinkles' front;
 \item Find (if it exists) the corresponding critical amount of shear such that 
 Eq.~\eqref{detM}
 is satisfied. 
\end{enumerate}

Then repeat Steps \emph{(ii)} and \emph{(iii)}
for other angles $\theta$ until the entire surface is spanned, 
and keep the smallest critical amount of shear $K_\text{cr}$ (say) 
for the angle $\Phi$ chosen in Step \emph{(i)}.
Then take a different value of 
$\Phi$, until all possible fibre orientations are covered.
\emph{In fine} a graph of $K_\text{cr}$ as a function of $\Phi$ is 
generated.


\section{Numerical results for biological soft tissues}


We take the \emph{standard reinforcing model}, 
\be \label{standard}
W = \mu(I_1 - 3)/2 + E(I_4-1)^2/4,
\en
where $E$ is an extensional modulus in the fibre direction.
This model has been used for several soft tissues, such as 
papillary muscle \cite{Tabe04}, myocardium  \cite{Tabe04},
skeletal muscles \cite{RoPu07}, or brainstem \cite{Ning06}.

\begin{figure}
\centering 
\epsfig{figure=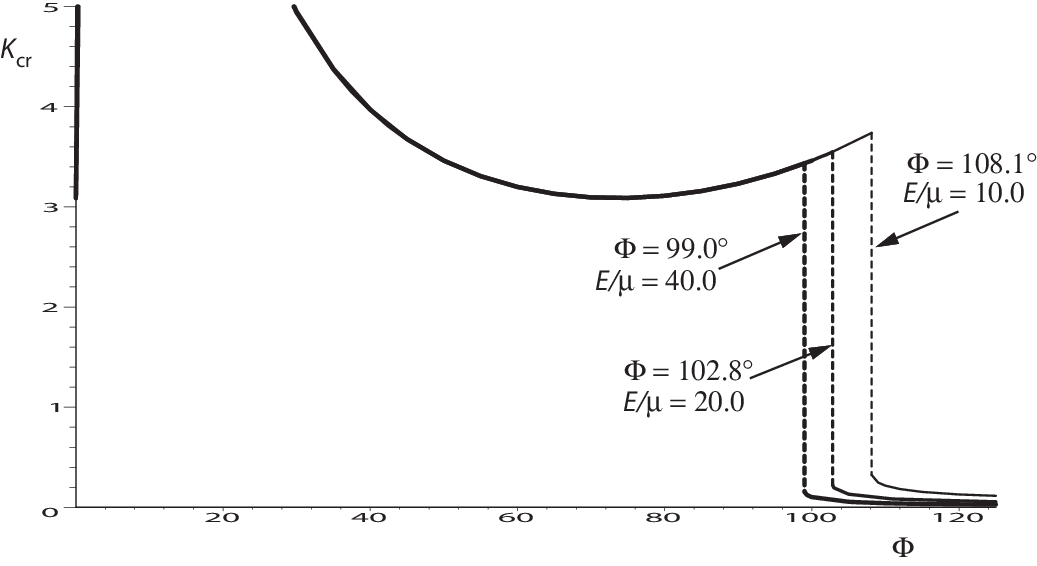, width=.78\textwidth, height=.42\textwidth}
 \caption{Variations of the critical amount of shear for surface
instability 
 with the angle between the directions of shear and the fibres. The solid 
 is modelled as a neo-Hookean matrix reinforced with one family 
 of fibres (standard reinforcing model);
 the ratio of the matrix shear modulus to the fibre stiffness is taken 
 in turn as $40.0$, $20.0$, and $10.0$.
 The 3 graphs coincide as long as $0 < \Phi < \Phi_0$, 
 where $\Phi_0 = 99.0^\circ$, $102.8^\circ$, $108.1^\circ$, respectively.
 At $\Phi \simeq \Phi_0$, the half-space switches from being very stable 
 ($K_\text{cr} > 3.09$) to being easily unstable ($K_\text{cr} < 0.3$).
 The part of the plot corresponding to $K_\text{cr} > 5$ is not shown 
 for physical and visual reasons.}
 \label{graph_one_family}
\end{figure}

That latter reference examines the ability of the 
constitutive model Eq.~\eqref{standard} to describe the mechanical response
of porcine brainstem specimens. 
Recall that large deformations, in particular large shears, of brain tissue are 
often associated with traumatic brain injuries (Doorly and Gilchrist, 2006).
Ning et al. \cite{Ning06} find that the model provides good agreement
with experimental data;
they estimate that for 4 week old pigs, $E$ is about 20 times larger than $\mu$.
In a recent review on physical properties of tissues for arterial ultrasound, 
Hoskins \cite{Hosk07} emphasizes the need for constitutive models of
nonlinear elastic 
behavior. 
He also collects available data for arterial walls: 
in particular for abdominal aortic aneurysms, ex vivo measurements indicate that 
$E$ is about 10 times larger than $\mu$ whilst for human atherosclerotic plaque, 
$E$ seems to be more than 40 times $\mu$.
For our numerical computations we take in turn the values 
$E/\mu = 40.0$, $20.0$, $10.0$, and collect the corresponding results on 
Fig.~\ref{graph_one_family}.

Broadly speaking, we find a region where the solid is strongly 
reinforced by the family of fibres, followed by an abrupt drop in the value of the 
critical amount of shear for surface instability, which occurs earlier as 
$E/\mu$ increases.

When the fibres are aligned with the direction of shear, they are not stretched and
they play no role;
thus it is appropriate that at  $\Phi = 0.0^\circ$, we find $K_\text{cr} = 3.09$,
the critical amount of shear for an isotropic neo-Hookean half-space, see Section 2. 

Next we find that $K_\text{cr}$ shoots up to unrealistic values when 
$\Phi \gtrsim 0.0^\circ$: for instance  $K_\text{cr} = 32.48$ 
when $\Phi = 3.0^\circ$ (not represented for visual convenience).
Hence, the solid is strongly reinforced with respect to surface stability when the shear 
takes place more or less along the fibres: wrinkling is prevented.

As the angle $\Phi$ between the shear and the fibres increases, the critical 
amount of shear goes through a maximum, then a minimum, always remaining above 
$3.09$, the value for an isotropic neo-Hookean half-space, as long as 
$\Phi \le \Phi_0$, where $\Phi_0 = 99.0^\circ$, $102.8^\circ$, $108.1^\circ$,
approximatively, for $E/\mu = 40.0, 20.0, 10.0$, respectively. 
It is worth noting that in the range $90.0^\circ < \Phi < \Phi_0$, 
the fibres undergo a slight compression at low shear levels, and then 
are in extension until the critical amount of shear is reached; 
even when the fibres are compressed, the half-space remains stable.

When the angle $\Phi$ is large, $\Phi_0 < \Phi < 180.0^\circ$,
the half-space becomes unstable at low amounts of shear. 
For instance at  $\Phi = 99.07^\circ$, we find that $K_\text{cr} = 0.153$
when $E/\mu = 40.0$;
note that in reaching that critical amount of shear, the fibres are compressed by 
less than $1.3\%$.
The switch from high  to low critical amounts of  shear is abrupt, 
due to the non-monotonicity of $\text{det } \vec{Z}$ with $K$: this
quantity has a 
minimum in the high range ($K > 3.09$) which is always negative
(indicating the existence of a root to  
Eq.~\eqref{detM}), but it can also have a minimum in the low range ($K <
0.3$). This minimum is positive 
when $\Phi < \Phi_0$ (no root to Eq.~\eqref{detM}) but negative when  $\Phi > \Phi_0$, hence the 
jump in $K_\text{cr}$.

Finally we note that in the range $\Phi_0 < \Phi < 180.0^\circ$, 
the angle $\theta$ normal to the wrinkles' 
front is close to $\Phi$ (within $2^\circ$), indicating that the wrinkles 
are almost at right-angle with the fibres;
these predictions are in accordance with the observation 
of Fig.~\ref{fig_silicone_meat}.


\section{Discussion}


We developed a quantitative methodology to understand the formation
of wrinkles in some biological soft tissues.
The analysis allowed us to model some visual observations of a
sheared elastomer versus a sheared piece of skeletal muscle, 
based on a simple nonlinear anisotropic
constitutive law (requiring the knowledge of only one quantity, $E/\mu$).

Studying the geometry and mechanics of wrinkles is relevant to many
biomechanical applications 
such as for instance the treatment of scars, and our results 
may provide some help in developing rational approaches to these
problems. 
The next logical step is to apply and generalize this methodology to
model the wrinkling of skin and other biological membranes.
These may require more work than here, with 
the consideration of two families of parallel fibres 
(the collagen network), 
but the methodology remains essentially the same.
It is also exact, versatile, and more convenient to apply than methods based on approximate 
theories (e.g. F\"oppl-von K\'arm\'an plate equations) because it
can accommodate easily 
anisotropy, nonlinear constitutive laws, finite thickness, 
and large homogeneous pre-deformation.



\end{document}